\documentclass[aps,prl,nobibnotes,reprint,longbibliography,floatfix]{revtex4-2}

\usepackage[utf8]{inputenc}
\usepackage[T1]{fontenc}
\usepackage{amsmath,amssymb,latexsym,graphicx}
\usepackage{newtxtext,newtxmath}
\usepackage{bbold,anyfontsize}
\usepackage[dvipsnames]{xcolor}
\usepackage[
  colorlinks=true,
  urlcolor=BlueViolet,
  citecolor=BlueViolet,
  filecolor=BlueViolet,
  linkcolor=BlueViolet
]{hyperref}

\begin{document}

\title{
  Algorithm-independent bounds on complex optimization through the statistics of marginal optima
}

\author{Jaron Kent-Dobias}
\email{jaron.kent-dobias@roma1.infn.it}
\affiliation{Istituto Nazionale di Fisica Nucleare, Sezione di Roma I, Rome, Italy 00184}

\begin{abstract}
  Optimization seeks extremal points in a function. When there are
  superextensively many optima, optimization algorithms are liable to get
  stuck. Under these conditions, generic algorithms tend to find marginal
  optima, which have many nearly flat directions. In a companion paper, we
  introduce a technique to count marginal optima in random landscapes. Here, we
  use the statistics of marginal optima calculated using this technique to
  produce generic bounds on optimization, based on the simple principle that
  algorithms will overwhelmingly tend to get stuck only where marginal optima are found. We
  demonstrate the idea on a simple non-Gaussian problem of maximizing the sum
  of squared random functions on a compact space. Numeric experiments using
  both gradient descent and generalized approximate message passing algorithms
  fall inside the expected bounds.
\end{abstract}

\maketitle

In optimization one usually looks for the global maximum or minimum of an
objective or energy function. Whether this is possible or likely depends on the
structure of the particular problem. Many optimization problems are
understood to have three phases: an easy phase, where a good global solution
can be quickly found by a well-chosen algorithm; an impossible phase, where no
such solution exists; and a hard phase, where a proliferation of bad local
solutions make the search for the good global one interminable \cite{Zdeborova_2016_Statistical, Gamarnik_2022_Disordered}.

In the hard phase, the bad local optima found by algorithms are not
completely random: in the large-problem limit, the same algorithm run many
times on many statistically identical systems finds optima with the same
properties. For a long time physicists believed that many algorithms would find
optima at a special energy level, the threshold energy $E_\mathrm{th}$, where
level sets of the energy transition from containing mostly saddles to
containing mostly minima. Recent work has shown that this is not actually true
\cite{Folena_2020_Rethinking, Folena_2021_Gradient, Folena_2023_On}. However, there is a kernel of
truth in the old idea.

As the transition point in energy between mostly saddles and mostly minima, the
threshold energy itself is characterized by mostly \emph{marginal minima},
which are minima whose Hessian has a pseudogap and that therefore have many nearly flat directions \cite{Muller_2015_Marginal}. While the threshold energy
is now understood to not be a meaningful attractor of dynamics, marginal minima
appear to be robust attractors of dynamics across a wide variety of models and algorithms.
There are multiple, complimentary ways to understand why this might be so. On
one hand, their flat directions make their basins of attraction naïvely much
larger than those of minima with only stiff directions. On the other, marginal
minima can gregariously accumulate along their flat directions, giving rise to
large flat regions that are likewise easy to find \cite{Baldassi_2016_Unreasonable, Baldassi_2021_Unveiling, Baldassi_2023_Typical}. However, these intuitions
cannot be understood to form a proof; rather, the attractiveness of marginal
minima is right now an empirical observation.

This empirical fact is now fairly well established among a variety of hard
models under a variety of algorithms \cite{Parisi_1995-01_On, Horner_2007_Time, Erba_2024_Quenches, Folena_2022_Marginal}. Gradient flow and descent, stochastic
gradient descent, Langevin or Monte Carlo annealing all find asymptotically
marginal minima in many complex landscapes. This is also true of approximate message
passing algorithms whose dynamics runs outside configuration space but
nonetheless end at marginal minima. The exceptions to this rule are few and
represent non-generic methods. For instance, stiff solutions to random
\textsc{xorsat} can be found by a nonlocal and non-generic change of variables that
renders the transformed problem easy \cite{Foini_2012_On}. And of course, any problem in its easy
phase finds the good solution whether it is stiff or not.

Taking the inevitability of marginal minima as a fact, one can try to derive
effective bounds on performance by inverting the logic: if generic algorithms in
complex settings find marginal optima, then the range of performances of generic
algorithms must lie in the range of energies in which marginal optima can be
found. This motivates the study of marginal optima and their properties
independent of our algorithm of interest. In our companion paper, we introduce
a general technique for studying the properties of marginal optima
\cite{Kent-Dobias_2024_Conditioning}. Here, we will use these results to test the
relationship between the distribution of marginal optima and the performance of
algorithms in a family of simple non-Gaussian energy landscapes.

Consider the problem of maximizing the sum of squared random functions
constrained to a manifold. In this case, the energy to minimize is minus the
sum of squares
\begin{equation}
  H(\mathbf x)=-\frac12\sum_{k=1}^{M}V_k(\mathbf x)^2
\end{equation}
of $M=\alpha N$ centered Gaussian random functions $V_k$. The configurations $\mathbf
x\in\mathbb R^N$ are constrained to lie on the hypersphere $\|\mathbf x\|^2=N$.
The functions $V_k$ are defined such that their covariance at two different points
$\mathbf x,\mathbf x'$ is
\begin{equation}
  \overline{V_i(\mathbf x)V_j(\mathbf x')}=N\delta_{ij}f\left(
    \frac{\mathbf x\cdot\mathbf x'}N
  \right)
\end{equation}
for some function $f$ with positive series coefficients. Each $V_k$ can be
understood as consisting of sums of fully connected, independently normally
distributed $p$-term interactions between the components of $\mathbf x$ whose
variance is proportional to $f^{(p)}(0)$, the $p$th series coefficient of $f$.

This problem has been studied recently because of its resemblance to other
problems in physics, mathematics, and optimization, like nonlinear least
squares and vertex models of tissues \cite{Fyodorov_2019_A, Fyodorov_2020_Counting,
Fyodorov_2022_Optimization, Tublin_2022_A, Vivo_2024_Random, Urbani_2023_A,
Kamali_2023_Stochastic, Kamali_2023_Dynamical, Urbani_2024_Statistical,
Montanari_2023_Solving, Montanari_2024_On}. However, those motivations favor
the opposite optimization problem of minimizing the sum of squares, or
equivalently maximizing $H$. Unfortunately minimizing the sum of squares in
this model is a poor platform to test the idea introduced in this paper. This
previous research suggests that for many choices of $f$ the problem of
minimizing the sum of squares is characterized by global and near-global optima
belonging to a wide flat manifold, the signature of so-called full replica
symmetry breaking (\textsc{rsb}) order. Not only are such cases more difficult
to treat analytically, but they also often \emph{easy} for certain algorithms
\cite{Gamarnik_2021-10_The}. The problem of maximizing the sum of squares, or
minimizing $H$, instead is typically characterized by isolated solutions belonging to the \emph{hard} family of {\oldstylenums1}\textsc{rsb}
problems. Motivated by our goal of producing bounds for performance in the hard
regime, we will study the relationship between marginal minima of $H$ and the
performance of algorithms.

In the physics of spin glasses there are a well-worn set of tools for studying
the statistics of optima in certain kinds of random hard problems. In such
problems, the number of optima $\mathcal N_H(E)$ at a certain energy level $E=H/N$ is
typically exponentially large in the size $N$ of the problem. The reliable
measure of typical statistics of optima is therefore the average of the
logarithm this number, which defines the complexity
$\Sigma(E)=\frac1N\overline{\log\mathcal N_H(E)}$. When the complexity $\Sigma$
is positive there are typically exponentially many optima, while when it is
negative it is extremely unlikely to find any.

Remarkably, this typical count
can be computed in a variety of fully-connected models, including by a method
which involves a Legendre transformation of a mere equilibrium calculation
\cite{Monasson_1995_Structural}.
A more flexible approach is offered by an old technique called the Kac--Rice method
\cite{Kac_1943_On, Rice_1939_The}. Here, the number of optima is found by integrating a Dirac $\delta$ function containing the gradient over configuration space, weighted by a Jacobian that consists of the Hessian:
\begin{equation} \label{eq:complexity.def}
  \mathcal N_H(E)=\int d\mathbf x\,\delta\big(\nabla H(\mathbf x)\big)\,
  \big|\det\operatorname{Hess}H(\mathbf x)\big|\,
  \delta\big(NE-H(\mathbf x)\big)
\end{equation}
The use of this method in the statistical physics of random landscapes is well
established in diverse settings \cite{Cavagna_1997_An, Muller_2006_Marginal,
Ros_2019_Complex, Kent-Dobias_2023_How}.

Unfortunately these traditional approaches are not sufficient for the purposes of
characterizing marginal optima. The typical complexity $\Sigma(E)$
corresponds with the statistics of marginal minima only at the threshold energy
$E_\mathrm{th}$. For smaller energies it characterizes stiff minima, while
for larger energies it characterizes unstable saddle points. To characterize the
typical complexity of marginal minima $\Sigma_\mathrm m(E)$, we need to
condition the count of stationary points on their marginality. Both
traditional approaches to calculating the complexity present unique problems.
In the Kac--Rice approach, conditioning on marginality requires a complicated
condition on the Hessian of the energy at a stationary point. In all but
Gaussian landscapes, properties of the Hessian at a point are dependent on its stationarity and energy \cite{Fyodorov_2004_Complexity, Bray_2007_Statistics}. Characterizing the Hessian at stationary
points in non-Gaussian problems is a class of calculation just in its infancy \cite{Maillard_2020_Landscape}.
Further conditioning the problem using the result of such a calculation is
further still. On the other hand, the Legendre transform technique does not
offer obvious ways to add further conditions on the properties of the
stationary points being counted.

In our companion paper, we introduce a method for calculating the marginal
complexity that avoids these problems by sidestepping the need to thoroughly understand the spectral problems of the conditioned Hessian \cite{Kent-Dobias_2024_Conditioning}. This method has two
steps: first, we insert in the integral \eqref{eq:complexity.def} a
resolution of a Dirac $\delta$ function that conditions the smallest eigenvalue
$\lambda_\mathrm{min}$ of the Hessian to take a specific value $\lambda^*$. Of
course, marginal minima correspond with $\lambda^*=0$, but the converse is not
necessarily true: spectra with isolated atoms at zero satisfy this but are not
marginal. Therefore, a second step is necessary to ensure the conditioned minima are marginal.

We further condition the count on the fact that
$\operatorname{Tr}\operatorname{Hess}H=N\mu$, i.e., that the trace of the
Hessian is fixed to a specific value. When $\mu$ is large and $\lambda^*=0$, the spectra of typical points
have a continuous bulk which is positive and an isolated
atom at zero. When $\mu$ is decreased far enough, the bulk of the spectrum will
hit zero and the large-deviation calculation that corresponds to fixing the lowest eigenvalue will break
down, producing imaginary predictions. By adjusting $\mu$ and looking for this
breakdown, we can tune the spectrum of stationary points to have a
pseudogap, and therefore be marginal, without actually understanding the form
of the bulk spectrum. At each energy we expect a different value $\mu_\mathrm m(E)$ of
the trace to reach this marginal point.

In the companion paper, we calculate the $\lambda^*$- and $\mu$-dependent
complexity of optima in the maximum of sum of squares problem under the assumption that its structure is {\oldstylenums1}\textsc{rsb}. The result takes the form of
an extremal problem
\begin{widetext}
\begin{equation} \label{eq:complexity}
  \begin{aligned}
    &\Sigma_{\lambda^*}(E,\mu)=\underset{\hat\beta,\hat\lambda,r,d,g,y,\Delta z}{\operatorname{extremum}}\Bigg[
    \hat\beta E-\mu(r+g+\hat\lambda)
    +\hat\lambda\lambda^*
    +\frac12\log\left(\frac{d+r^2}{g^2}\times\frac{y^2-2\Delta z}{y^2}\right)
    \\
    &-\frac\alpha2\log\Bigg(
      \frac{
        1-2(2y+\hat\lambda)f'(1)+4(y^2-2\Delta z)f'(1)^2
      }{\big[1-2yf'(1)\big]^2}
      \times\frac{
        f(1)\big[f'(1)d-\hat\beta-f''(1)(r^2-g^2+8(y\hat\lambda+\Delta z)+2\hat\lambda^2)\big]+\big[1-rf'(1)\big]^2
      }{
        \big[1+gf'(1)\big]^2
      }
  \Bigg)\Bigg]
  \end{aligned}
\end{equation}
\end{widetext}
In problems like the one considered here, which have all noise and no signal,
there is another simplification. Under these circumstances, typical minima do not
have isolated eigenvalues, and pulling one out of the continuous part of the spectrum always decreases the
complexity. Therefore, the complexity resulting from fixing the minimum eigenvalue
$\lambda^*$ to be at the bottom edge of the bulk spectrum will be stationary
with respect to the position of the eigenvalue, or
$0=\frac\partial{\partial\lambda^*}\Sigma_{\lambda^*}(E,\mu)$. We can thus
choose the marginal trace $\mu_\mathrm m(E)$ as the value of $\mu$ such that the complexity is
stationary with respect to variation of the minimum eigenvalue about zero, or by solving
\begin{equation} \label{eq:marginal.trace}
  0=\frac\partial{\partial\lambda^*}\Sigma_{\lambda^*}\big(E,\mu_\mathrm m(E)\big)\bigg|_{\lambda^*=0}.
\end{equation}
for $\mu_\mathrm m(E)$.
This allows us to define the marginal complexity $\Sigma_\mathrm m(E)=\Sigma_0(E,\mu_\mathrm m(E))$ in a purely variational way.

\begin{figure}
  \includegraphics{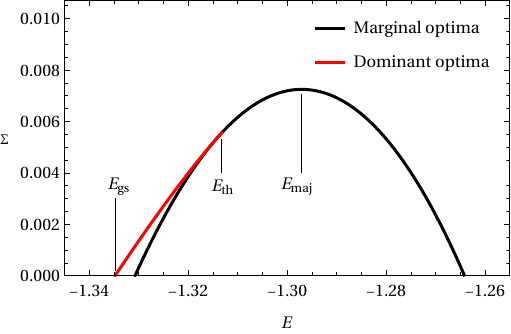}
  \caption{
    The marginal and dominant complexities of the random sum of squares problem
    with $\alpha=\frac18$ and covariance function $f(q)=\frac14q+\frac34q^2$. Several important
    energies are marked on the plot: the ground state energy $E_\mathrm{gs}$
    where the lowest optima are found, the threshold energy $E_\mathrm{th}$
    where marginal optima are the most common kind of optima, and the majority
    energy $E_\mathrm{maj}$ where most marginal optima are found.
  } \label{fig:marginal.complexity}
\end{figure}

Combining the $\lambda^*$-dependent complexity \eqref{eq:complexity} with the
requirement \eqref{eq:marginal.trace} allows us to compute the marginal
complexity for the problem of maximizing sums of squared random functions. This is plotted as a function of energy $E$ in comparison to the
dominant complexity for a representative example of the maximum random sum of
squares problem in Fig.~\ref{fig:marginal.complexity}. That figure shows
several features that are ubiquitous in optimizing hard complex problems.
Indeed, they are ubiquitous enough features that this plot appears
schematically almost identical to that of \citeauthor{Folena_2020_Rethinking}
for the Gaussian spherical spin glasses, despite relying on a more
sophisticated calculation \cite{Folena_2020_Rethinking}. First, marginal minima are found in exponential number at energies far above the threshold energy, which a naïve
analysis of the dominant complexity might lead one to conclude is the highest
energy at which minima can be found. Second, there is a gap in energy between the
lowest marginal minima and the ground state. This is a gulf that one cannot
hope to cross with a generic algorithm.

How does the performance of real algorithms compare with these broad bounds? We
studied this question by performing two prototypical algorithms for complex
optimization: gradient descent and generalized approximate message passing. The
first, gradient descent, starts at a random point in configuration space and
takes strictly downward steps in the direction of the gradient whose magnitude
is determined by a simple line search. The second, generalized approximate
message passing, was introduced for the spherical spin glasses under another
name and in a slightly different form by \citeauthor{Subag_2020_Following} \cite{Subag_2020_Following, ElAlaoui_2021_Optimization}. This algorithm starts outside of
configuration space with a unit vector in a random direction. Every step is a
unit vector in the direction of the gradient but orthogonal to the current
state. After $N$ such steps, the configuration lies on the radius $\sqrt N$
hypersphere. In order to ensure that the resulting configuration is a minimum,
we afterwards run gradient descent using this configuration as the starting
condition. In the Gaussian spherical spin glasses this algorithm has been
proven to be asymptotically optimal among a broad class of generic algorithms \cite{ElAlaoui_2021_Optimization}.
In the present non-Gaussian problem of maximizing random sums of squares it is
not known whether it is optimal, but our experiments indicate that it usually outperforms gradient descent.

\begin{figure}
  \includegraphics{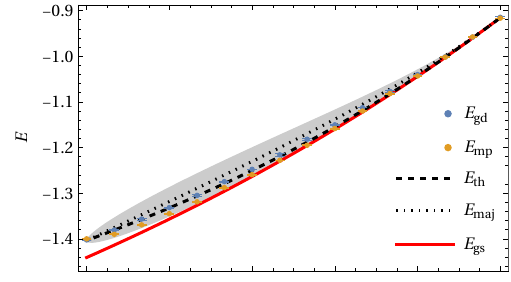}
  \includegraphics{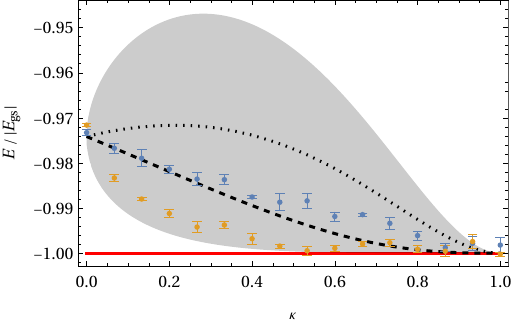}
  \caption{
    Comparison of the marginal complexity with the asymptotic performance of
    gradient descent ($E_\mathrm{gd}$) and generalized approximate message
    passing algorithms ($E_\mathrm{mp}$) for maximum sum of squares models with $\alpha=\frac18$ and
    covariance function $f(q)=\kappa q+(1-\kappa)q^2$ as $\kappa$ is varied.
    The shaded region shows the range of energies at which the marginal
    complexity is positive. The lines indicated on the legend show the
    threshold energy, majority energy, and ground state energy. The bottom plot
    is a rescaling of the top plot by the ground state energy to make
    the comparison with marginal complexity easier. Error bars show one
    standard error.
  } \label{fig:alg.test}
\end{figure}

We ran these algorithms on independent random instances of the maximum sum of
squares problem with $\alpha=\frac18$ and covariance functions $f(q)=\kappa q+(1-\kappa)q^2$ for a
variety of $\kappa$ between zero and one and for a variety of system sizes
between $N=16$ and $2048$. First, we confirmed that both algorithms indeed find
asymptotically marginal minima by calculating the average of the smallest
eigenvalue of the Hessian $\lambda_\textrm{min}(N)$ as a function of system
size. As expected, this smallest eigenvalue asymptotically approaches zero as
$N$ becomes large, and we measured power law scaling. We further found that the energy achieved asymptotically approaches its infinite-size value with a power law whose exponent depends on $\kappa$. Details of our
numeric analysis can be found in the supplementary material.

The asymptotic results of these numeric experiments are compared with the
predictions of the marginal complexity in Fig.~\ref{fig:alg.test}. The range of
energies where marginal complexity is positive succeed at effectively bounding
the algorithmic performance over the entire range of models, though often the bound
is not very tight. The threshold energy $E_\mathrm{th}$ is often not
asymptotically reached by gradient descent and is usually beaten by message
passing; as for the spherical spin glasses, it doesn't have a straightforward
relationship with algorithmic performance, though it is possibly a lower bound
for gradient descent from a random initial condition \cite{Folena_2023_On}. In
this family of models and for the spherical spin glasses, the energy
$E_\textrm{maj}$ at which the majority of marginal minima are found may be an
effective upper bound on the performance of gradient descent, but we do not
have a principled reason for this.

In this paper, we have demonstrated on a non-Gaussian complex optimization
problem that the complexity of marginal minima can be used to effectively bound
algorithmic performance independent of the algorithm used. The computation of
this marginal complexity in the non-Gaussian case was made possible by the
techniques introduced in our companion paper. When comparing with algorithmic
performance in a `worst case' (gradient descent) and `best case' (generalized
approximate message passing), the bounds set by marginal complexity are
followed, albeit often loosely.

Despite being loose, the bounds derived by marginal complexity represent a
useful development. The method used to compute the marginal complexity
introduced in our companion paper is general should can be applied to
other systems in much the same way as equilibrium calculations, despite being
more complex than an equilibrium calculation.  The current rigorous and often tight
lower bounds on performance developed for spin glass models using the overlap
gap property have yet to be generalized to non-Gaussian continuous models \cite{Gamarnik_2021-01_The, Gamarnik_2021-10_The, Huang_2022_Computational}. It is also an
open question, even in simple spherical spin glasses, at what energy the `worst
case' of gradient flow arrives. This means that marginal complexity can make a
initial, perhaps loose, lower \emph{and} upper bound on behavior without
needing to develop substantially new model-specific approaches.

Some future development of the method is necessary for efficiently applying it
to systems with a nonzero signal. In such cases, there are circumstances where
typical stationary points have an isolated eigenvalue, and the variational
approach of \eqref{eq:marginal.trace} for finding the shift of the
spectrum corresponding to a pseudogap will fail \cite{Ros_2019_Complex}. While we
described a generic method for doing this by looking for the breakdown of
the large-deviation principle behind the calculation, an extended variational
approach would be much easier to work with. It is possible something of this
kind could be accomplished by treating the signal direction and the directions
orthogonal to it separately.

The numeric comparisons made in this paper were constrained to small powers of
$q$ in $f(q)$ and to small $\alpha$ by the computational feasibility of getting
sufficient statistics from the simulations. With use of the dynamical mean
field theory results already developed for the sum of squared random functions,
comparisons with asymptotic dynamics should be made for a wider range of model
parameters \cite{Kamali_2023_Dynamical, Kamali_2023_Stochastic}. Further development of the marginal complexity in the full
\textsc{rsb} setting for comparison with the least squares problem should also
be made.

In order to improve the bounds made by marginal complexity, it would be beneficial
to understand if there is something structural about the type of marginal
minima that can attract some algorithmic dynamics and those that cannot
attracted any. Recent analysis of the two-point complexity of marginal minima
in the spherical models did not find such a signature
\cite{Kent-Dobias_2024_Arrangement}. It was recently proposed that traits related to
the self-similarity and stochastic stability of minima might be important, but these
ideas are still in their infancy \cite{Urbani_2024_Statistical}.

\begin{acknowledgements}
  JK-D is supported by a \textsc{DynSysMath} Specific Initiative of the INFN.
  The authors thank James P Sethna, Ralph B Robinson, and Bethany Dixon at
  Cornell University for providing and facilitating access to computing
  resources used in this work.
\end{acknowledgements}

\bibliography{marginal}

\end{document}


\title{
  \emph{Supplement to}\\Algorithm-independent bounds on complex optimization through the statistics of marginal optima
}

\author{Jaron Kent-Dobias}
\affiliation{Istituto Nazionale di Fisica Nucleare, Sezione di Roma I, Rome, Italy 00184}

\begin{abstract}
  This supplement describes in detail the procedure and analysis of the numeric
  experiments whose results appear in the main text.
\end{abstract}

\maketitle

\section{Implementation of the energy}

Define the projection matrix
\begin{equation}
  P_\mathrm x=I-\frac{\mathbf x\mathbf x^T}{\|\mathbf x\|^2}
\end{equation}
which projects a vector into the subspace orthogonal to $\mathbf x$. Then the
gradient can be derived from the derivative with respect to $\mathbf x$ by
\begin{equation}
  \nabla H(\mathbf x)=P_\mathbf x\partial H(\mathbf x)
\end{equation}
Likewise the Hessian matrix is
\begin{equation}
  \operatorname{Hess} H(\mathbf x)
  =P_\mathbf x\left[\partial\partial H(\mathbf x)-\frac{\mathbf x^T\partial H(\mathbf x)}NI\right]P_\mathbf x^T
\end{equation}
The Hessian so defined has a zero eigenvalue whose corresponding eigenvector is
$\mathbf x$. This must be discarded in any analysis of the stability.

\section{Specification of the algorithms}

Our gradient descent algorithm is standard, with a simple line search to
determine the step size whose acceptance is made with the Wolfe criterion
\cite{Wolfe_1969_Convergence}. The algorithm is halted when the squared norm of
the gradient falls under a certain threshold $\epsilon N$, where we take
$\epsilon=10^{-13}$. The choice for $\epsilon$ was made because it is roughly
the smallest value possible before some loops hang due to machine imprecision.

\begin{algorithm}[H]
  \caption{Gradient descent}
  \label{alg:grad}
  \begin{algorithmic}
  \Function{Normalize}{$\mathbf x$}
    \State\Return $\frac{\sqrt N}{\|\mathbf x\|}\mathbf x$
  \EndFunction
  \State $\mathbf x \gets \mathbf x_0$
  \State $\alpha \gets 1$
  \While{$\|\nabla H(\mathbf x)\|^2 / N > \epsilon$}
    \While{$\mathbf x'\gets\Call{Normalize}{\mathbf x-\alpha\nabla H(\mathbf x)}$,\\
      $H(\mathbf x') > H(\mathbf x)-\frac12\alpha\|\nabla H(\mathbf x)\|^2$}
      \State $\alpha \gets \frac12\alpha$
    \EndWhile
    \State $\mathbf x\gets\mathbf x'$
    \State $\alpha \gets \frac54\alpha$
  \EndWhile
  \State\Return $\mathbf x$
  \end{algorithmic}
\end{algorithm}

In its original form, the algorithm introduced by
\citeauthor{Subag_2020_Following} used steps proportional to an eigenvector of
the Hessian \cite{Subag_2020_Following}. However, equivalent asymptotic results have been proven to result
from steps along the direction of the gradient. Note that based on our
definition of the gradient, the steps are orthogonal to the state $\sigma$ of
the algorithm by construction.

\begin{algorithm}[H]
  \caption{Approximate message passing}
  \label{alg:mp}
  \begin{algorithmic}
  \State $\pmb\sigma \gets \mathbf x_0 / N$
  \While{$\|\pmb\sigma\|^2<N$}
    \State $\pmb\sigma\gets\pmb\sigma - \nabla H(\pmb\sigma) / \|\nabla H(\pmb\sigma)\|$
  \EndWhile
  \State\Return $\pmb\sigma$
  \end{algorithmic}
\end{algorithm}

After the \textsc{amp} algorithm has terminated we run gradient descent using
its output as the initial condition to ensure that the final result is a
minimum. The code implementing these algorithms is freely available \cite{Kent-Dobias_2024_least-squares}.

\section{Experiments and raw data}

Our numeric experiments consisted of running these algorithms many times on
independent realizations of the disorder, and measuring the final energy and
the minimum eigenvalue of the Hessian. We made extensive use of GNU Parallel
command line tool \cite{Tange_2011_GNU}. Linear algebra manipulations were all
done using the Eigen library. Experiments were run for models with
$\alpha=\frac18$ and $f(q)=\kappa q+(1-\kappa)q^2$ at 16 values of $\kappa$
evenly spaced between $0$ and $1$ inclusive, and for system sizes $N$ at powers
of 2 from 16 to 2048. The number of independent experiments at each set of parameters is given in Table~\ref{tab:N}.

Here we share the summary statistics for the raw data in the form of the mean
and standard error of the mean. Tables \ref{tab:Eg} and \ref{tab:Em} show these
statistics for the final energies of gradient descent and \textsc{amp},
respectively, while Tables \ref{tab:λg} and \ref{tab:λm} show the same
statistics for the minimum eigenvalue of the Hessian for gradient descent and
\textsc{amp}, respectively.

\section{Analysis}

\subsection{Confirmation that marginal minima are found}

\begin{figure}
  \includegraphics{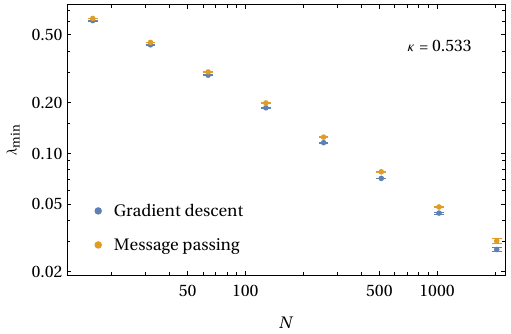}
  \caption{
    Log--log plot of the minimum eigenvalue $\lambda_\mathrm{min}$ as a
    function of $N$ measured in our experiments for $\kappa=0.533$.
  } \label{fig:eigenvalue}
\end{figure}

The behavior of the minimum eigenvalue in each case is consistent with a power
law in $N$ asymptotically approaching zero. This is evidenced for a
representative set of data from $\kappa=0.533$ in Fig.~\ref{fig:eigenvalue}.
We concluded from this data that the points found by our algorithms are indeed
asymptotically marginal.

\subsection{Fitting the asymptotic energy}

The results in the main text are extrapolations of the raw measurements above.
Since good extrapolation is a tricky problem, we share in detail our procedure
for estimating the asymptotic results.

\begin{figure}
  \includegraphics{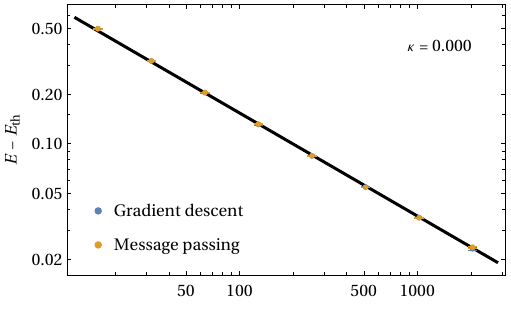}

  \vspace{-1em}

  \includegraphics{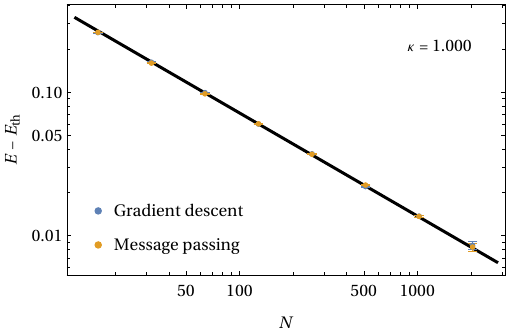}
  \caption{
    The approach of algorithmic performance to the mean-field predicted
    threshold energy of the model for $\kappa=0.000$ and $\kappa=1.000$, where
    marginal minima are predicted to exist at only one energy. The black lines
    show power-law fits with exponents $\gamma_0=0.624\pm0.004$ and
    $\gamma_1=0.718\pm0.007$, respectively.
  }\label{fig:k_convergence}
\end{figure}

First, the mean-field analysis predicts that at $\kappa=0$ and $\kappa=1$,
marginal minima only appear at a single energy. Therefore, both algorithms
should also approach that energy. Fig.~\ref{fig:k_convergence} shows the
approach of the measured performance of both algorithms to the predicted
threshold energy in these two cases, and the results are consistent with
power-law approaches that are the same for the two algorithms but different at
different $\kappa$. The measured exponents are  $\gamma_0=0.624\pm0.004$ and
$\gamma_1=0.718\pm0.007$ for $\kappa=0$ and $\kappa=1$, respectively.

\begin{figure}[htbp]
  \includegraphics{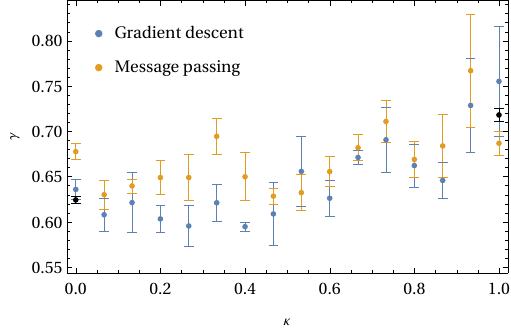}
  \caption{
    The fit value of the exponent $\gamma$ for the two algorithms at all values
    of $\kappa$. The black points for $\kappa=0$ and $1$ are the results of our
    fit of the exponent in Fig.~\ref{fig:k_convergence} where the asymptotic
    value of the energy was an input in the fit.
  } \label{fig:gamma}
\end{figure}

We therefore have a reasonable expectation that the convergence of the energy
of each algorithm to its asymptotic value should be well-described by a power
law depending on $\kappa$. We fit to each set of data the function
\begin{equation}
  E(N)=E_\mathrm{alg}+(a_0+a_1N+a_2N^2)^{-\gamma/2}
\end{equation}
which is a truncated expansion in large $N$ about the asymptotic value. The
values of the exponent $\gamma$ resulting from this fit can be found in
Fig.~\ref{fig:gamma}, where they are also compared with the fit at $\kappa=0$
and 1 that used knowledge of the asymptotic energy. The values of the exponent
predicted by these fits is roughly consistent with their more precise value at
these two points.

\begin{figure}[htbp]
  \includegraphics{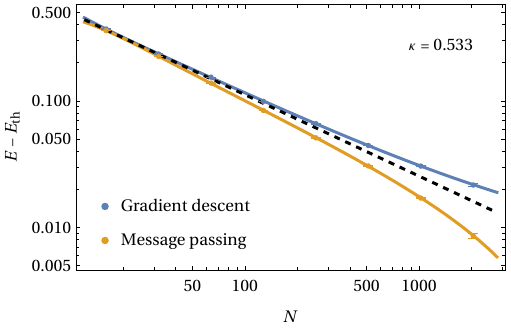}

  \caption{
    The approach of algorithmic performance to the mean-field predicted
    threshold energy of the model for $\kappa=0.533$. The solid lines show our
    fit to the data, while the dashed line shows a pure power-law whose
    exponent is the same as the fits.
  } \label{fig:mid}
\end{figure}

Table~\ref{tab:analysis} shows the values of the energy that were plotted in
the figure in the main text, both from the mean-field prediction and from these
asymptotic fits. For another perspective on the relationship between our data,
fits, and the mean-field predictions, Fig.~\ref{fig:mid} shows the approach of
the algorithmic performance to the mean-field threshold at an intermediate
value $\kappa=0.533$. There, neither algorithm appears to asymptotically
approach the threshold energy, instead approaching energies above and below.

\bibliography{marginal}

\clearpage

\begin{table*}
  \begin{tabular}{|l||r|r|r|r|r|r|r|r|}
    \hline
    $\kappa$ & $N=16$ & $N=32$ & $N=64$ & $N=128$ & $N=256$ & $N=512$ & $N=1024$ & $N=2048$\\\hline
0.000 & 10000 & 10000 & 10000 & 10000 & 10000 & 10000 & 2712 & 296\\ 
0.067 & 10000 & 10000 & 10000 & 10000 & 10000 & 10000 & 1706 & 227\\ 
0.133 & 10000 & 10000 & 10000 & 10000 & 10000 & 10000 & 1799 & 242\\ 
0.200 & 10000 & 10000 & 10000 & 10000 & 10000 & 10000 & 1820 & 254\\ 
0.267 & 10000 & 10000 & 10000 & 10000 & 10000 & 10000 & 1897 & 262\\ 
0.333 & 10000 & 10000 & 10000 & 10000 & 10000 & 10000 & 1990 & 274\\ 
0.400 & 10000 & 10000 & 10000 & 10000 & 10000 & 10000 & 2026 & 284\\ 
0.467 & 10000 & 10000 & 10000 & 10000 & 10000 & 10000 & 2132 & 291\\ 
0.533 & 10000 & 10000 & 10000 & 10000 & 10000 & 10000 & 2140 & 266\\ 
0.600 & 10000 & 10000 & 10000 & 10000 & 10000 & 10000 & 2270 & 295\\ 
0.667 & 10000 & 10000 & 10000 & 10000 & 10000 & 10000 & 2390 & 297\\ 
0.733 & 10000 & 10000 & 10000 & 10000 & 10000 & 10000 & 1898 & 319\\ 
0.800 & 10000 & 10000 & 10000 & 10000 & 10000 & 10000 & 1994 & 365\\ 
0.867 & 10000 & 10000 & 10000 & 10000 & 10000 & 10000 & 2390 & 296\\ 
0.933 & 10000 & 10000 & 10000 & 10000 & 10000 & 10000 & 2669 & 342\\ 
1.000 & 10000 & 10000 & 10000 & 10000 & 10000 & 10000 & 3570 & 256\\\hline
  \end{tabular}
  \caption{
    Number of independent samples in each of the measurements reported here.
  } \label{tab:N}
\end{table*}

\begin{table*}
  \begin{tabular}{|l||l|l|l|l|l|l|l|l|}
    \hline
    $\kappa$ & $N=16$ & $N=32$ & $N=64$ & $N=128$ & $N=256$ & $N=512$ & $N=1024$ & $N=2048$\\\hline
0.000 & $-0.90798(173)$ & $-1.08681(119)$ & $-1.20043(78)$ & $-1.27367(49)$ & $-1.31994(31)$ & $-1.34949(19)$ & $-1.36840(23)$ & $-1.38103(42)$\\ 
0.067 & $-0.89882(173)$ & $-1.07332(120)$ & $-1.18424(79)$ & $-1.25510(51)$ & $-1.29939(32)$ & $-1.32750(19)$ & $-1.34606(29)$ & $-1.35870(55)$\\ 
0.133 & $-0.89681(176)$ & $-1.06173(120)$ & $-1.16618(80)$ & $-1.23453(51)$ & $-1.27791(32)$ & $-1.30519(20)$ & $-1.32412(30)$ & $-1.33497(52)$\\ 
0.200 & $-0.88798(177)$ & $-1.04550(122)$ & $-1.14865(80)$ & $-1.21303(51)$ & $-1.25469(33)$ & $-1.28126(21)$ & $-1.29835(31)$ & $-1.31045(52)$\\ 
0.267 & $-0.87660(180)$ & $-1.02661(121)$ & $-1.12894(80)$ & $-1.19104(52)$ & $-1.23022(33)$ & $-1.25646(21)$ & $-1.27275(30)$ & $-1.28363(56)$\\ 
0.333 & $-0.86618(179)$ & $-1.01193(123)$ & $-1.10826(82)$ & $-1.16761(52)$ & $-1.20546(34)$ & $-1.23027(21)$ & $-1.24608(30)$ & $-1.25565(55)$\\ 
0.400 & $-0.85336(179)$ & $-0.99313(122)$ & $-1.08574(81)$ & $-1.14308(53)$ & $-1.17929(33)$ & $-1.20294(22)$ & $-1.21826(30)$ & $-1.22829(54)$\\ 
0.467 & $-0.83896(180)$ & $-0.97360(124)$ & $-1.06183(82)$ & $-1.11717(52)$ & $-1.15093(34)$ & $-1.17439(21)$ & $-1.18867(31)$ & $-1.19803(50)$\\ 
0.533 & $-0.82028(179)$ & $-0.95490(121)$ & $-1.03533(80)$ & $-1.09002(52)$ & $-1.12292(33)$ & $-1.14447(21)$ & $-1.15830(30)$ & $-1.16714(52)$\\ 
0.600 & $-0.80336(180)$ & $-0.92861(122)$ & $-1.01045(81)$ & $-1.06132(52)$ & $-1.09318(33)$ & $-1.11376(21)$ & $-1.12646(29)$ & $-1.13555(55)$\\ 
0.667 & $-0.78234(178)$ & $-0.90690(122)$ & $-0.98288(80)$ & $-1.03123(52)$ & $-1.06153(33)$ & $-1.08069(21)$ & $-1.09269(27)$ & $-1.10042(46)$\\ 
0.733 & $-0.76018(180)$ & $-0.87895(120)$ & $-0.95269(80)$ & $-0.99954(51)$ & $-1.02951(33)$ & $-1.04739(21)$ & $-1.05858(31)$ & $-1.06641(48)$\\ 
0.800 & $-0.73753(179)$ & $-0.85193(122)$ & $-0.92277(80)$ & $-0.96739(51)$ & $-0.99464(33)$ & $-1.01148(21)$ & $-1.02201(30)$ & $-1.02932(47)$\\ 
0.867 & $-0.71079(177)$ & $-0.81968(120)$ & $-0.89046(80)$ & $-0.93342(51)$ & $-0.95877(32)$ & $-0.97430(20)$ & $-0.98450(26)$ & $-0.99067(45)$\\ 
0.933 & $-0.68189(179)$ & $-0.79114(121)$ & $-0.85529(78)$ & $-0.89637(51)$ & $-0.92134(32)$ & $-0.93553(20)$ & $-0.94471(25)$ & $-0.95055(41)$\\ 
1.000 & $-0.65664(179)$ & $-0.75501(121)$ & $-0.81712(80)$ & $-0.85626(52)$ & $-0.87901(32)$ & $-0.89415(21)$ & $-0.90243(22)$ & $-0.90752(51)$\\\hline
  \end{tabular}
  \caption{
    Measured data for the energy $E_\mathrm{gd}$ achieved by gradient descent
    optimization from a random initial condition on the maximum sum-of-squares
    problem with $f(q)=\kappa q+(1-\kappa)q^2$ and $\alpha=\frac18$. Each column contains the mean
    value of many experiments on independent disorder, while the parentheses
    give the standard uncertainty of the mean in the last digits.
  } \label{tab:Eg}
\end{table*}

\begin{table*}
  \begin{tabular}{|l||l|l|l|l|l|l|l|l|}
    \hline
    $\kappa$ & $N=16$ & $N=32$ & $N=64$ & $N=128$ & $N=256$ & $N=512$ & $N=1024$ & $N=2048$\\\hline
0.000 & $-0.90656(173)$ & $-1.08690(120)$ & $-1.19982(78)$ & $-1.27261(49)$ & $-1.31950(31)$ & $-1.34970(19)$ & $-1.36852(22)$ & $-1.38044(39)$\\ 
0.067 & $-0.91469(172)$ & $-1.08552(116)$ & $-1.19829(77)$ & $-1.26681(48)$ & $-1.31157(29)$ & $-1.33963(18)$ & $-1.35771(25)$ & $-1.36920(42)$\\ 
0.133 & $-0.91405(174)$ & $-1.07836(117)$ & $-1.18447(76)$ & $-1.25167(46)$ & $-1.29430(29)$ & $-1.32138(17)$ & $-1.33857(24)$ & $-1.35000(41)$\\ 
0.200 & $-0.91057(173)$ & $-1.06693(117)$ & $-1.16777(73)$ & $-1.23270(46)$ & $-1.27352(28)$ & $-1.29962(17)$ & $-1.31580(24)$ & $-1.32705(38)$\\ 
0.267 & $-0.89521(172)$ & $-1.05252(116)$ & $-1.14785(75)$ & $-1.21030(46)$ & $-1.25002(29)$ & $-1.27478(17)$ & $-1.29076(23)$ & $-1.30160(36)$\\ 
0.333 & $-0.88474(171)$ & $-1.03517(115)$ & $-1.12680(74)$ & $-1.18710(46)$ & $-1.22476(28)$ & $-1.24881(17)$ & $-1.26356(23)$ & $-1.27324(38)$\\ 
0.400 & $-0.86774(173)$ & $-1.01317(116)$ & $-1.10325(75)$ & $-1.16232(46)$ & $-1.19740(28)$ & $-1.22014(17)$ & $-1.23465(23)$ & $-1.24375(36)$\\ 
0.467 & $-0.85316(170)$ & $-0.98973(115)$ & $-1.07897(74)$ & $-1.13361(47)$ & $-1.16786(28)$ & $-1.18933(17)$ & $-1.20336(22)$ & $-1.21201(36)$\\ 
0.533 & $-0.83058(170)$ & $-0.96488(117)$ & $-1.05088(75)$ & $-1.10509(46)$ & $-1.13785(28)$ & $-1.15823(17)$ & $-1.17169(22)$ & $-1.18014(36)$\\ 
0.600 & $-0.81277(173)$ & $-0.93960(116)$ & $-1.02260(74)$ & $-1.07325(46)$ & $-1.10527(29)$ & $-1.12487(17)$ & $-1.13713(22)$ & $-1.14542(36)$\\ 
0.667 & $-0.78874(175)$ & $-0.91342(117)$ & $-0.99298(75)$ & $-1.04123(47)$ & $-1.07146(29)$ & $-1.08986(18)$ & $-1.10157(22)$ & $-1.10844(35)$\\ 
0.733 & $-0.76714(176)$ & $-0.88650(118)$ & $-0.96093(77)$ & $-1.00723(47)$ & $-1.03656(29)$ & $-1.05406(18)$ & $-1.06479(25)$ & $-1.07188(35)$\\ 
0.800 & $-0.73947(179)$ & $-0.85407(117)$ & $-0.92827(77)$ & $-0.97263(48)$ & $-0.99904(30)$ & $-1.01608(19)$ & $-1.02627(24)$ & $-1.03263(35)$\\ 
0.867 & $-0.70943(174)$ & $-0.82420(119)$ & $-0.89440(77)$ & $-0.93470(48)$ & $-0.96119(30)$ & $-0.97666(19)$ & $-0.98654(24)$ & $-0.99250(39)$\\ 
0.933 & $-0.68125(178)$ & $-0.79198(119)$ & $-0.85592(78)$ & $-0.89678(50)$ & $-0.92221(31)$ & $-0.93639(20)$ & $-0.94523(23)$ & $-0.95106(37)$\\ 
1.000 & $-0.65594(183)$ & $-0.75720(122)$ & $-0.81898(80)$ & $-0.85608(51)$ & $-0.87925(32)$ & $-0.89360(21)$ & $-0.90247(22)$ & $-0.90775(55)$\\\hline
  \end{tabular}
  \caption{
    Measured data for the energy $E_\mathrm{mp}$ achieved by approximate message passing
    optimization from a random initial condition on the maximum sum-of-squares
    problem with $f(q)=\kappa q+(1-\kappa)q^2$ and $\alpha=\frac18$. Each column contains the mean
    value of many experiments on independent disorder, while the parentheses
    give the standard uncertainty of the mean in the last digits.
  } \label{tab:Em}
\end{table*}

\begin{table*}
  \begin{tabular}{|l||l|l|l|l|l|l|l|l|}
    \hline
    $\kappa$ & $N=16$ & $N=32$ & $N=64$ & $N=128$ & $N=256$ & $N=512$ & $N=1024$ & $N=2048$\\\hline
0.000 & $0.71936(421)$ & $0.51324(281)$ & $0.33927(181)$ & $0.21732(115)$ & $0.13387(70)$ & $0.08259(44)$ & $0.05023(51)$ & $0.03129(93)$\\ 
0.067 & $0.70286(404)$ & $0.50420(272)$ & $0.33152(176)$ & $0.21167(112)$ & $0.13254(69)$ & $0.08093(42)$ & $0.04881(62)$ & $0.02991(104)$\\ 
0.133 & $0.69834(400)$ & $0.49053(265)$ & $0.32481(173)$ & $0.20872(109)$ & $0.13008(68)$ & $0.07968(42)$ & $0.04958(62)$ & $0.03187(106)$\\ 
0.200 & $0.68035(390)$ & $0.48309(260)$ & $0.31915(167)$ & $0.20267(107)$ & $0.12623(65)$ & $0.07833(40)$ & $0.04735(58)$ & $0.02957(99)$\\ 
0.267 & $0.66124(375)$ & $0.46599(250)$ & $0.31318(165)$ & $0.19987(104)$ & $0.12447(65)$ & $0.07774(41)$ & $0.04772(56)$ & $0.02979(94)$\\ 
0.333 & $0.65032(368)$ & $0.45774(245)$ & $0.30733(162)$ & $0.19671(102)$ & $0.12219(64)$ & $0.07548(39)$ & $0.04559(52)$ & $0.02763(90)$\\ 
0.400 & $0.63585(355)$ & $0.45019(241)$ & $0.30042(156)$ & $0.19179(100)$ & $0.12026(63)$ & $0.07400(38)$ & $0.04523(52)$ & $0.02768(86)$\\ 
0.467 & $0.62176(353)$ & $0.44066(235)$ & $0.29470(154)$ & $0.18781(98)$ & $0.11631(61)$ & $0.07244(38)$ & $0.04560(51)$ & $0.02669(79)$\\ 
0.533 & $0.60421(333)$ & $0.43434(227)$ & $0.28738(149)$ & $0.18442(95)$ & $0.11472(59)$ & $0.07059(36)$ & $0.04402(48)$ & $0.02682(84)$\\ 
0.600 & $0.59116(332)$ & $0.42272(221)$ & $0.28278(147)$ & $0.18163(93)$ & $0.11263(58)$ & $0.07013(36)$ & $0.04386(47)$ & $0.02706(79)$\\ 
0.667 & $0.57549(315)$ & $0.41720(218)$ & $0.27889(142)$ & $0.17834(92)$ & $0.11116(57)$ & $0.06851(36)$ & $0.04251(44)$ & $0.02557(76)$\\ 
0.733 & $0.56843(315)$ & $0.40670(209)$ & $0.27253(140)$ & $0.17377(90)$ & $0.10984(56)$ & $0.06851(35)$ & $0.04148(49)$ & $0.02482(67)$\\ 
0.800 & $0.56446(303)$ & $0.40596(211)$ & $0.26942(138)$ & $0.17280(89)$ & $0.11005(57)$ & $0.06794(35)$ & $0.04234(49)$ & $0.02542(71)$\\ 
0.867 & $0.56650(299)$ & $0.40422(209)$ & $0.26579(139)$ & $0.17300(89)$ & $0.10909(56)$ & $0.06820(35)$ & $0.04289(44)$ & $0.02695(75)$\\ 
0.933 & $0.57989(310)$ & $0.40681(214)$ & $0.26445(138)$ & $0.17036(88)$ & $0.10888(56)$ & $0.06789(34)$ & $0.04303(42)$ & $0.02703(71)$\\ 
1.000 & $0.62006(347)$ & $0.40045(221)$ & $0.25863(144)$ & $0.16637(92)$ & $0.10520(58)$ & $0.06789(37)$ & $0.04245(39)$ & $0.02644(92)$\\\hline
  \end{tabular}
  \caption{
    Measured data for the minimum eigenvalue $\lambda_\mathrm{min,gd}$ of the
    Hessian matrix at the optimum found by gradient descent optimization from a
    random initial condition on the maximum sum-of-squares problem with
    $f(q)=\kappa q+(1-\kappa)q^2$ and $\alpha=\frac18$. Each column contains the mean value of many
    experiments on independent disorder, while the parentheses give the
    standard uncertainty of the mean in the last digits.
  } \label{tab:λg}
\end{table*}

\begin{table*}
  \begin{tabular}{|l||l|l|l|l|l|l|l|l|}
    \hline
    $\kappa$ & $N=16$ & $N=32$ & $N=64$ & $N=128$ & $N=256$ & $N=512$ & $N=1024$ & $N=2048$\\\hline
0.000 & $0.71938(417)$ & $0.51263(279)$ & $0.34091(182)$ & $0.21405(113)$ & $0.13347(70)$ & $0.08284(43)$ & $0.05028(50)$ & $0.02905(85)$\\ 
0.067 & $0.71827(409)$ & $0.51024(272)$ & $0.34111(179)$ & $0.21664(113)$ & $0.13767(71)$ & $0.08433(44)$ & $0.05112(62)$ & $0.03199(115)$\\ 
0.133 & $0.72265(410)$ & $0.51023(270)$ & $0.33633(174)$ & $0.21749(110)$ & $0.13659(70)$ & $0.08440(43)$ & $0.05343(64)$ & $0.03070(108)$\\ 
0.200 & $0.70819(393)$ & $0.50095(265)$ & $0.33180(170)$ & $0.21504(109)$ & $0.13350(68)$ & $0.08465(43)$ & $0.05173(61)$ & $0.03275(110)$\\ 
0.267 & $0.68157(380)$ & $0.49125(257)$ & $0.32417(166)$ & $0.21083(106)$ & $0.13159(66)$ & $0.08284(41)$ & $0.05149(59)$ & $0.03264(106)$\\ 
0.333 & $0.68096(372)$ & $0.48421(251)$ & $0.32048(165)$ & $0.20799(104)$ & $0.13008(65)$ & $0.08147(41)$ & $0.05059(56)$ & $0.03199(98)$\\ 
0.400 & $0.65990(362)$ & $0.47288(244)$ & $0.31322(161)$ & $0.20207(102)$ & $0.12839(65)$ & $0.08071(40)$ & $0.05020(54)$ & $0.02974(89)$\\ 
0.467 & $0.64214(344)$ & $0.45833(236)$ & $0.30726(153)$ & $0.19883(100)$ & $0.12532(63)$ & $0.07867(39)$ & $0.04781(51)$ & $0.03064(91)$\\ 
0.533 & $0.62187(333)$ & $0.44892(227)$ & $0.30034(151)$ & $0.19685(97)$ & $0.12373(60)$ & $0.07717(38)$ & $0.04781(51)$ & $0.03011(90)$\\ 
0.600 & $0.60851(322)$ & $0.43718(223)$ & $0.29416(148)$ & $0.19112(94)$ & $0.12063(59)$ & $0.07575(37)$ & $0.04748(49)$ & $0.03010(83)$\\ 
0.667 & $0.59020(316)$ & $0.42813(215)$ & $0.29062(144)$ & $0.18506(91)$ & $0.11833(58)$ & $0.07433(36)$ & $0.04598(44)$ & $0.02876(82)$\\ 
0.733 & $0.58186(311)$ & $0.42092(214)$ & $0.28442(140)$ & $0.18296(89)$ & $0.11722(57)$ & $0.07245(36)$ & $0.04536(50)$ & $0.02778(77)$\\ 
0.800 & $0.57098(302)$ & $0.41448(208)$ & $0.27944(139)$ & $0.18020(89)$ & $0.11374(55)$ & $0.07243(35)$ & $0.04496(49)$ & $0.02855(72)$\\ 
0.867 & $0.56885(293)$ & $0.41109(209)$ & $0.27692(138)$ & $0.17709(86)$ & $0.11344(55)$ & $0.07123(34)$ & $0.04510(45)$ & $0.02903(81)$\\ 
0.933 & $0.58620(308)$ & $0.41033(211)$ & $0.26824(135)$ & $0.17506(86)$ & $0.11218(55)$ & $0.07045(35)$ & $0.04371(40)$ & $0.02813(70)$\\ 
1.000 & $0.62062(347)$ & $0.40689(224)$ & $0.26055(145)$ & $0.16625(92)$ & $0.10557(58)$ & $0.06708(36)$ & $0.04262(38)$ & $0.02829(102)$\\\hline
  \end{tabular}
  \caption{
    Measured data for the minimum eigenvalue $\lambda_\mathrm{min,mp}$ of the
    Hessian matrix at the optimum found by approximate message passing optimization from a
    random initial condition on the maximum sum-of-squares problem with
    $f(q)=\kappa q+(1-\kappa)q^2$ and $\alpha=\frac18$. Each column contains the mean value of many
    experiments on independent disorder, while the parentheses give the
    standard uncertainty of the mean in the last digits.
  } \label{tab:λm}
\end{table*}

\begin{table*}
  \begin{tabular}{|l||l|l|l|l|l|l|l|}
    \hline
    $\kappa$ & $E_\mathrm{max}$ & $E_\mathrm{maj}$ & $E_\mathrm{gd}$ & $E_\mathrm{th}$ & $E_\mathrm{mp}$ & $E_\mathrm{min}$ & $E_\mathrm{gs}$ \\ \hline
0.000 & $-1.4040575$ & $-1.4040575$ & $-1.4028(9)$ & $-1.4040575$ & $-1.4004(6)$ & $-1.4040575$ & $-1.4413071$ \\ 
0.067 & $-1.3534309$ & $-1.3756461$ & $-1.3812(16)$ & $-1.3811572$ & $-1.3905(12)$ & $-1.3982616$ & $-1.4140567$ \\ 
0.133 & $-1.3188326$ & $-1.3475583$ & $-1.3571(26)$ & $-1.3576074$ & $-1.3697(6)$ & $-1.3767883$ & $-1.3862994$ \\ 
0.200 & $-1.2868229$ & $-1.3187015$ & $-1.3319(13)$ & $-1.3326088$ & $-1.3452(12)$ & $-1.3513189$ & $-1.3571483$ \\ 
0.267 & $-1.2565960$ & $-1.2895687$ & $-1.3052(19)$ & $-1.3064171$ & $-1.3193(15)$ & $-1.3234285$ & $-1.3269697$ \\ 
0.333 & $-1.2279759$ & $-1.2606161$ & $-1.2751(15)$ & $-1.2793533$ & $-1.2881(10)$ & $-1.2940793$ & $-1.2961899$ \\ 
0.400 & $-1.1997200$ & $-1.2307909$ & $-1.2481(4)$ & $-1.2504947$ & $-1.2598(15)$ & $-1.2626306$ & $-1.2638278$ \\ 
0.467 & $-1.1719637$ & $-1.2004436$ & $-1.2164(24)$ & $-1.2201357$ & $-1.2284(5)$ & $-1.2296515$ & $-1.2302922$ \\ 
0.533 & $-1.1448189$ & $-1.1700419$ & $-1.1822(21)$ & $-1.1886609$ & $-1.1954(11)$ & $-1.1957413$ & $-1.1960622$ \\ 
0.600 & $-1.1171307$ & $-1.1383232$ & $-1.1506(12)$ & $-1.1550201$ & $-1.1588(8)$ & $-1.1599089$ & $-1.1600522$ \\ 
0.667 & $-1.0888977$ & $-1.1056071$ & $-1.1131(3)$ & $-1.1195931$ & $-1.1204(6)$ & $-1.1226685$ & $-1.1227239$ \\ 
0.733 & $-1.0600370$ & $-1.0722906$ & $-1.0775(14)$ & $-1.0828870$ & $-1.0821(8)$ & $-1.0846003$ & $-1.0846179$ \\ 
0.800 & $-1.0289907$ & $-1.0367316$ & $-1.0405(10)$ & $-1.0437575$ & $-1.0436(8)$ & $-1.0445305$ & $-1.0445345$ \\ 
0.867 & $-0.9952971$ & $-0.9990875$ & $-1.0018(9)$ & $-1.0027533$ & $-1.0026(13)$ & $-1.0029953$ & $-1.0029958$ \\ 
0.933 & $-0.9584445$ & $-0.9595729$ & $-0.9585(15)$ & $-0.9605787$ & $-0.9581(15)$ & $-0.9606113$ & $-0.9606114$ \\ 
1.000 & $-0.9160000$ & $-0.9160000$ & $-0.9151(15)$ & $-0.9160000$ & $-0.9169(4)$ & $-0.9160000$ & $-0.9160000$\\\hline
  \end{tabular}
  \caption{
    Significant energy values plotted in the main text. Numbers without
    uncertainty are the result of a mean-field calculation and have arbitrary
    precision. Numbers with uncertainty are the result of an extrapolation of
    data to infinite $N$ described in the text of this supplement.
  } \label{tab:analysis}
\end{table*}